\documentclass[prl,twocolumn,amsmath,amssymb,showpacs,superscriptaddress]{revtex4}

\usepackage{graphicx,epstopdf,bm}
\usepackage{subfigure}
\usepackage{soul}
\usepackage{hyperref}
\hypersetup{colorlinks=true,linkcolor=black,citecolor=black,urlcolor=black}
\usepackage[normalem]{ulem}
\usepackage{color} 
   
\usepackage{amsmath}
\usepackage[export]{adjustbox}

\usepackage{sidecap}
\sidecaptionvpos{figure}{c}  

\newcommand{\Fkt}[1]{\,\mathsf {#1}}
\ifx\Tr\renewcommand{\Tr}{\Fkt{Tr}}
\else\newcommand{\Tr}{\Fkt{Tr}}
\fi

\begin{document}

\title{Weak-Field Coherent Control of Ultrafast Molecule Making}


\author{Moran Geva}
\affiliation{The Shirlee Jacobs Femtosecond Laser Research Laboratory, Schulich Faculty of Chemistry, Technion-Israel Institute of Technology, Haifa 32000, Israel}

\author{Yonathan Langbeheim}
\affiliation{The Shirlee Jacobs Femtosecond Laser Research Laboratory, Schulich Faculty of Chemistry, Technion-Israel Institute of Technology, Haifa 32000, Israel}

\author{Arie Landau}
\affiliation{Institute of Advanced Studies in Theoretical Chemistry and Schulich Faculty of Chemistry, Technion-Israel Institute of Technology, Haifa 32000, Israel}

\author{Zohar Amitay}
\thanks{E-mail: amitayz@technion.ac.il}
\affiliation{The Shirlee Jacobs Femtosecond Laser Research Laboratory, Schulich Faculty of Chemistry, Technion-Israel Institute of Technology, Haifa 32000, Israel}

 

\begin{abstract}

Coherent control of ultrafast molecule making from colliding reactants is crucial 
for realizing coherent control of binary photoreactions (CCBP). 
To handle diverse excitation scenarios, feasibility with both weak and strong fields is essential.
We experimentally demonstrate here the weak-field feasibility, achieving  
it even under thermally hot conditions typical of chemical reactions.
The making of KAr molecules from hot pairs of colliding K and Ar atoms via resonance-mediated two-photon excitation is 
controlled by weak linearly-chirped femtosecond pulses. 
Negative chirps enhance the yield.
Our experimental and {\it ab initio} 
results are in excellent agreement.
New routes to CCBP are opened. 
\looseness=-1


\end{abstract}


\maketitle 

\everypar{\looseness=-1}

One of the holy grails in the field of quantum coherent control 
is the yet-to-be-realized coherent control of 
binary chemical photoreactions~\cite{tannor1,RonnieDancing89,RiceBook,ShapiroBook}. 
The operating principle  is 
to use shaped femtosecond pulses~\cite{WeinerRevSciInst00} to control and actively drive
binary chemical reaction along its full desired path from the 
initial state of the reactants to the final state of the desired products. 
This is actually one of the dreams that led, more than thirty years ago, to the birth of the coherent control concept.
Its realization at high temperatures, which are typical of chemical reactions, will enable a novel type of photochemistry.
The first part of the scheme is a coherently controlled ultrafast making of a molecule from the 
pair of colliding reactants via its photo-excitation at short internuclear distances 
to target electronically excited molecular states.
Subsequently, in the second part, 
the generated molecule undergoes further photo-control, 
leading to its dissociation into the intended products.

The main unique challenges of such coherent control of ultrafast molecule making (CC-UMM) 
with thermal reactants 
~\cite{DantusCPL95,RybakFaraday11,RybakPRL11,AmaranJCP13,LevinPRL15,LevinJPhysB15,LevinJPhysB21},
particularly at high temperatures, 
are posed by the initial 
state of the system 
due to the incoherent population of a vast 
number of scattering eigenstates.
Since this situation is highly disadvantageous for coherent control, 
one challenge is the need 
to incorporate in the controlled photo-excitation a filtering mechanism that selects, 
out of the full initial incoherent ensemble of all the possible excitation channels,
a sub-ensemble of channels that are susceptible to coherent control.
This challenge is fundamental from both a theoretical and an experimental point of view.
Another challenge, of experimental nature, 
is the very low signal resulting from
the very small probability and yield existing here for the photo-excitation at short internuclear distances.
Obviously, overcoming the latter challenge to obtain a measurable signal is a prerequisite for experimentally addressing the former one.

The specific excitation scenario 
of the ultrafast molecule making 
is set by the involved molecular states and the transitions among them that are photo-induced by the laser pulse. 
Since different intensity regimes are best suited for coherently controlling different scenarios, 
in order to handle diverse excitation situations, feasibility with both weak and strong fields is essential.
So far, following many years 
without any experimental realization of CC-UMM 
despite its importance, 
CC-UMM was successfully demonstrated experimentally only in the strong-field regime 
with pairs of thermally-hot colliding atoms~\cite{LevinPRL15,LevinJPhysB15,LevinJPhysB21}.
Strong shaped femtosecond pulses controlled there 
the making of Mg$_{2}$ molecules 
from hot pairs of 
Mg atoms 
via a 
non-resonant two-photon transition followed by multiple Raman transitions.   
Here, we experimentally extend the CC-UMM feasibility to the 
weak-field regime. 
We coherently control by weak linearly-chirped femtosecond pulses
the making of KAr molecules from thermally-hot pairs of colliding K and Ar atoms via 
a resonance-mediated two-photon transition.
Our {\it ab initio} studies are in excellent agreement with~the experiments 
and explain them.
All this establishes another important milestone toward 
photoreaction coherent control.  

Since its introduction over twenty years ago, 
weak-field coherent control by shaped femtosecond pulses~(WCC-SFP)~\cite{
SilberbergNat98,
SilberbergPRA99,
ZamithPRL01,
SilberbergPRL01-ResMed1plus1,
StaufferJCP02,
SilberbergPRL02-CohTransient,
SilberbergNat02-Raman,
DegertPRL02,
SilberbergPRL02-Raman,
DantusJCP03,
ChatelPRA03-ResMed1plus1-chirp,
SilberbergPRL04-ResMed1plus1-PolShaping,
ChatelPRAA04-ResMed1plus1-chirp,
PrakeltPRA04,
LeonePRA04,
LeonePRA05-Raman,
GandmanPRA07-1,
GandmanPRA07-2,
AmitayPRL08,
GandmanPRL14}
has proven to be a highly effective and successful method for controlling few-photon excitations of initially-bound systems, 
both conceptually and practically.
The weak-field regime corresponds to a photo-excitation that is fully described 
within the corresponding lowest-order time-dependent perturbation theory.
For example, for a weak two-photon excitation, it is the second-order one. 
Such perturbative description is very powerful  
since it 
allows to fully identify the initial-to-final multiphoton  
pathways  
and understand the coherent interferences  
among them.
The control is then achieved by 
manipulating these interferences via the femtosecond pulse shaping. 
Despite its success, 
WCC-SFP has not yet been experimentally demonstrated with any initially-unbound system. 
So, the current work 
is highly significant also in terms of 
this general aspect, as the first one to accomplish it. 

\begin{figure}[t]
\includegraphics[width=0.95\columnwidth,left]{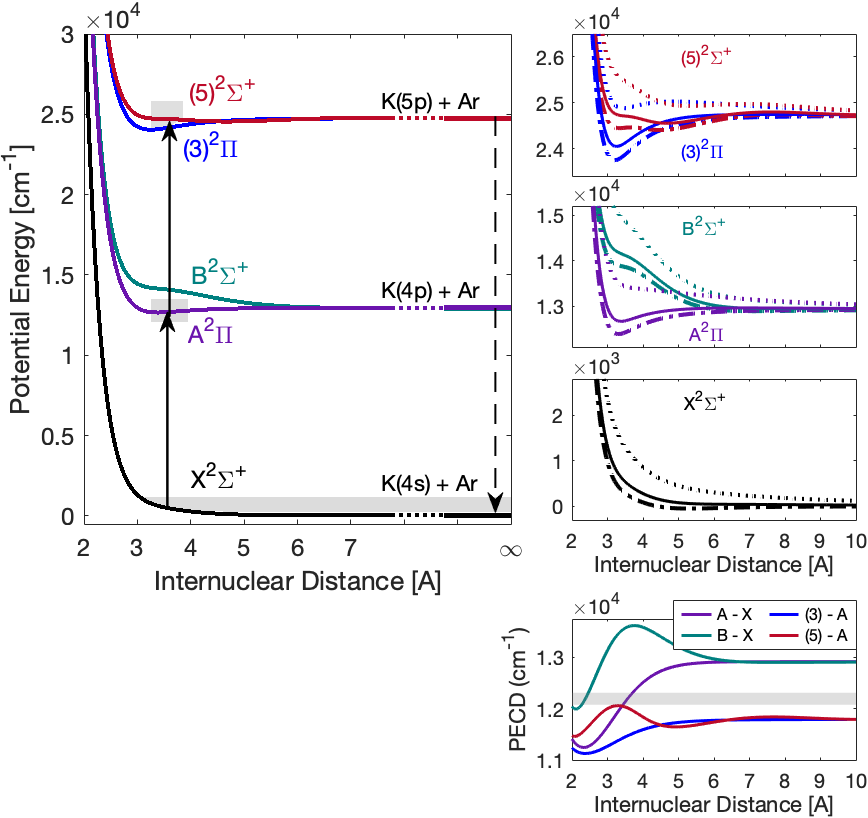}
\caption{
Excitation scheme and relevant potentials  
for the ultrafast making of a KAr molecule from a thermally-hot 
pair of colliding K and Ar atoms 
via a weak-field femtosecond resonance-mediated two-photon excitation. 
All the potentials are weakly bound or completely unbound, depending on the rotational quantum number J.
The right panels show the rotationless potentials (dash-dotted lines) and 
the examples of  
J=60 (solid lines) and 120 (dotted lines).
The left panel corresponds to J=60.
The shaded areas on the potentials show the thermally populated scattering states in X$^{2}\Sigma^{+}$
and the energetic ranges accessible via their excitation.
Relevant potential differences (PECDs) are also shown. 
The shaded area on the PECDs corresponds to the spectrum of the exciting pulse.
The generated molecules are detected upon their dissociation and the subsequent 
decay of their K($5p$) fragment via spontaneous emission. 
See text for details.
}
\label{fig:pot_exc_scheme}
\end{figure}

Figure~\ref{fig:pot_exc_scheme} shows the present excitation scheme of the 
ultrafast molecule making K+Ar+2$h\nu$~$\rightarrow$~KAr$^{*}$
with the relevant KAr molecular electronic states. 
Initially, a thermally-hot ensemble of colliding K and Ar atoms
populates the van-der-Waals ground electronic state X$^{2}\Sigma^{+}$ 
at a temperature of $T$=573~K ($k_{_B}T$$\approx$400~cm$^{-1}$).  
The X$^{2}\Sigma^{+}$ state has a very shallow rotationless well of only about 40~cm$^{-1}$. 
With the rotational-barrier addition, it does not support any bound levels for rotations (partial waves) above J$\approx$40. 
Hence, the initial thermal population of X$^{2}\Sigma^{+}$   
predominantly occupies (unbound)  scattering eigenstates. 
%
%
From these states, K-Ar collision pairs are photo-excited at short internuclear distances by a linearly-chirped femtosecond pulse 
in a molecular weak-field resonance-mediated two-photon excitation  
via the A$^{2}\Pi$ and B$^{2}\Sigma^{+}$ states to the $(3)^{2}\Pi$ and $(5)^{2}\Sigma^{+}$ states.
The pulse is of 820-nm central wavelength, $\sim$8.1-nm ($\sim$120-cm$^{-1}$) bandwidth, 
125-fs transform-limited (TL) duration, 5$\times$10$^{10}$-W/cm$^{2}$ TL peak intensity, and linear polarization.
It is not resonant with any one- or two-photon atomic transition of the individual colliding atoms.
Since  
the pulse spectrum is resonant 
only with the   
X$^{2}\Sigma^{+}$$-$A$^{2}\Pi$ and A$^{2}\Pi$$-$$(5)^{2}\Sigma^{+}$ potential differences (see Fig.~\ref{fig:pot_exc_scheme}),
the dominant electronic excitation route  
is X$^{2}\Sigma^{+}$$-$A$^{2}\Pi$$-$$(5)^{2}\Sigma^{+}$.
This identification is 
confirmed by the full theoretical results. 
In the following, we thus consider only this route. 
Similar to the X-state, the A$^{2}\Pi$ and $(5)^{2}\Sigma^{+}$ states are also weakly bound or completely unbound, depending on J.
Their rotationless well depths are of about 500 and 300~cm$^{-1}$ 
and their bound levels exist only up to J$\approx$90 and 80, respectively. 
Following the broad pulse spectrum, the excitation from each initial thermally-populated 
unbound eigenstate in X$^{2}\Sigma^{+}$ 
coherently goes through multiple intermediate eigenstates in A$^{2}\Pi$   
and ends up at multiple final eigenstates in $(5)^{2}\Sigma^{+}$. 
Both bound and unbound eigenstates are involved as the intermediates and finals. 
Hence, the entire controlled excitation 
is simultaneously composed of a thermal mixture of a vast number of channels, 
each starts from a different initial eigenstate. 
There is an extremely large variability in the excitation frequencies among the different channels.
The KAr molecules are generated at the $(5)^{2}\Sigma^{+}$ state with their nuclear radial 
distribution localized at short internuclear distances.
The population excited from bound eigenstates in X$^{2}\Sigma^{+}$ is negligible 
due to their insignificant 
initial population 
and very low excitation probability. 
The latter results from their excitation being non-resonant with the A$^{2}\Pi$ state, 
for which the longest corresponding resonant excitation wavelength is 794.9~nm~\cite{BokelmannJCP96_KAr_exp1,DurenJCP81_KAr_exp3}.
This negligible contribution is 
consistent with the experiments (see below) 
and confirmed by the full theoretical results.

The experiments take place
in a static cell at 573~K holding a gas mixture of potassium and argon,  
with the K vapor pressure being about 0.3~Torr and the Ar pressure set to 60~Torr.
The sample is irradiated at 1-kHz repetition rate by the linearly-chirped femtosecond pulses, 
after they undergo shaping in a setup incorporating a liquid-crystal spatial light phase modulator~\cite{WeinerRevSciInst00}.
Their spectral phase 
is of the form $\Phi(\omega) = \frac{1}{2} k (\omega - \omega_{0})^2$, 
where $\omega_{0}$ is the central frequency and $k$ is the linear chirp parameter.

\begin{SCfigure*}
\includegraphics[width=0.545\textwidth]{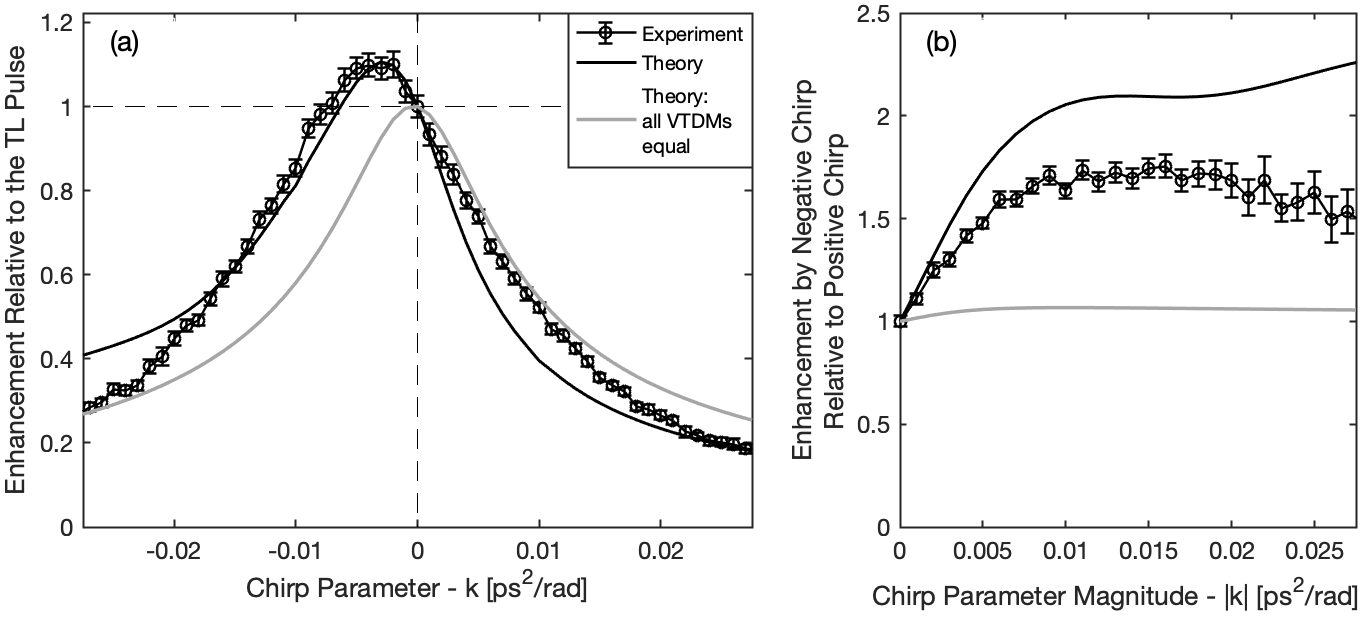} \hspace*{1mm}
\caption{Experimental and theoretical chirp-dependent results for the emission intensity $I_{5p}$.
(a) Enhancement by a linearly-chirped pulse relative to the TL pulse.
(b) Enhancement by a negatively-chirped pulse relative to the positively-chirped pulse of the same chirp magnitude.
The theoretical results include full results (black lines) and 
results with artificially setting all the system's vibrationally-averaged transition dipole moments (VTDMs) 
to the same value for eliminating the Franck-Condon filtering (gray lines).}
\label{fig:exp_theo_results}
\end{SCfigure*}

The KAr molecules formed at the $(5)^{2}\Sigma^{+}$ state 
are detected 
upon their dissociation to the atomic fragments K($5p$) and ground-state Ar 
at the state's asymptote, 
and the subsequent spontaneous decay of the K($5p$) via the $5p$$\rightarrow$$4s$  
transition. 
Due to the experimental conditions, the dissociating molecules include not only  
the molecules generated as unbound (which naturally dissociate), 
but also those generated as bound. The bound ones dissociate upon their collisions with the Ar atoms.
Based on experiments with other weakly-bound 
systems~\cite{ThompsonJCP82}, 
we calculated the present Ar-induced collisional dissociation lifetime of different $(5)^{2}\Sigma^{+}$'s bound levels to be shorter than 10~ns.
Such collisional dissociation dominants over 
the other decay channel of spontaneous emission, 
for which we have calculated a much longer lifetime of about 120-150~ns (depending on the level).
This radiative lifetime value is consistent with the $\sim$135-ns value measured for K($5p$)~\cite{BerendsACTAB88,MillsPRA05,SafronovaPRA08}.
There is no other source here for excited K($5p$) atoms, 
since the atomic two-photon transition between ground-state K($4s$) and excited K($5p$) is forbidden
as is the X$^{2}\Sigma^{+}$-to-$(5)^{2}\Sigma^{+}$ KAr two-photon transition at internuclear distances beyond $\sim$10~\AA.
So, the intensity of the K($5p$$\rightarrow$$4s$) radiation emitted from the excited sample 
is a background-free measure for the yield of KAr molecules formed at the $(5)^{2}\Sigma^{+}$ state.
Experimentally, the fluorescence from the cell is optically collected and then 
detected by a spectrometer and a time-gated camera system,
measuring the integrated intensity $I_{5p}$ of the 
K($5p$$\rightarrow$$4s$)  emission line 
at $\sim$404.5~nm. 
At our working conditions, 
as expected for a signal originating only from the excitation of K-Ar pairs, 
the $I_{5p}$ signal 
from a given pulse
exhibits linear dependence on each of the K and Ar partial pressures down to zero pressures, 
with no signal when either is zero. 
Also, as expected for a two-photon process in the weak-field regime,
the $I_{5p}$ signal 
from a given pulse exhibits quadratic dependence on the TL peak intensity down to zero intensity, 
while the 
chirp dependence of the $I_{5p}$ signal exhibits no dependence on~it.

The experimental results 
of weak-field coherent control of ultrafast molecule making 
are shown in Fig.~\ref{fig:exp_theo_results}. 
Figure~\ref{fig:exp_theo_results}(a) presents the measured $I_{5p}$, 
normalized with respect to the one obtained for the TL pulse ($k$=0),  
versus the chirp parameter $k$.
The presented quantity is thus the enhancement factor $\mathsf{EF}(k)$=$I_{5p}(k) / I_{5p}(k$=$0)$.
Several main features are observed for the chirp control.
One feature is the enhancement that negatively chirped pulses with small chirps  
exhibit over the TL pulse. 
The maximal $\mathsf{EF}$ of 
$\sim$1.10 is observed for chirps in the range of 
$-$0.002 to $-$0.005~ps$^2$/rad.
Another feature 
is the enhancement that any negatively chirped pulse 
exhibits over the positively chirped pulse 
of the same chirp magnitude. 
The corresponding enhancement factor, 
$\mathsf{EF_{N/P}}(|k|)$=$I_{5p}(-|k|) / I_{5p}(+|k|)$, 
is presented in Fig.~\ref{fig:exp_theo_results}(b) versus the chirp parameter magnitude ($|k|$).
Its value monotonically increases 
from 1 for $|k|$=0 to exceed 1.45
for all the present $|k|$ values 
above 0.005~ps$^2$/rad, with a weak chirp dependence there, 
reaching $\sim$1.75 within the range of $|k|$$\approx$0.01-0.02~ps$^2$/rad.
The degree of coherent control exhibited here is best reflected by 
$\mathsf{EF_{N/P}}$ 
since it corresponds to a pure phase effect, as the two corresponding 
chirped pulses have identical intensity but different phase in both the time and frequency domains.
In terms of the general chirp-dependence trend, 
for the negative chirps below the chirps of maximal $\mathsf{EF}$
as well as for all the positive chirps,
the $\mathsf{EF}$ continuously decreases as the chirp magnitude increases.
The observed enhancement by negative chirps also confirms that the contribution to the measured signal
from the bound eigenstates in X$^{2}\Sigma^{+}$ is non-significant. 
This is because their two-photon excitation is non-resonant with A$^{2}\Pi$ 
and thus leads to a symmetric signal with no such enhancement, 
i.e., it has equal values for negative and positive chirps of equal magnitude and maximal value for the TL pulse~\cite{DantusJCP03}.

To explain the experimental results, our theoretical studies consist of two parts. 
The first theoretical part includes 
state-of-the-art {\it ab initio} electronic structure calculations
of the KAr potential energy curves 
and electronic transition dipole moments (ETDMs). 
The calculations are performed using the non-relativistic (without spin-orbit coupling) quantum-chemistry package Q-Chem~\cite{QChem}.
We employ the electron-attachment equation-of-motion coupled-cluster method restricted to single and double excitations (EA-EOM-CCSD) 
with the aug-pc-3 basis set and frozen core.
The computation high accuracy is confirmed by the good agreement of the calculated spectroscopic constants 
with available experimental values for the X, A and B states~\cite{BokelmannJCP96_KAr_exp1,DurenZPA68_KAr_exp2,DurenJCP81_KAr_exp3,DurenJCP88_KAr_exp4,FalkePRA06_K4p_level,JohanssonPhySc72_K5p_level,NIST_ASD}.
The present {\it ab initio} approach for calculating the extended  
manifold of potentials is much more accurate than previous corresponding calculation~\cite{RhoumaJCP02}.
The full set of the ETDMs is calculated here for the first time.
The calculated potentials are those presented in Fig.~\ref{fig:pot_exc_scheme}.

The second theoretical part numerically calculates the shaped femtosecond excitation
using the frequency-domain picture obtained within second-order time-dependent perturbation theory.
The corresponding formulation is given in the 
Supplemental Material~\footnote{See Supplemental Material.}. 
For the nuclear dynamics, the calculations consider both the vibrational (radial) and rotational degrees of freedom.
The results of the electronic calculations are used to get the bound and unbound 
eigenstates of the different electronic states (for different values of J)   
and the various vibrationally-averaged transition dipole moments (VTDMs).
The eigenstates are numerically calculated using a computational box with a radius of 120~${\rm \AA}$.
The calculated signal $I_{5p}$ is taken as the total population of $(5)^{2}\Sigma^{+}$. 
To reduce the computational effort, the calculations only include 
excitations with equal initial and final rotations, i.e., $J_{\rm X}$=$J_{5\Sigma}$, 
instead of all the transition-allowed 
possibilities of $J_{5\Sigma}$$-$$J_{\rm X}$=0,$\pm$1,$\pm$2. 
The considered intermediates 
in A$^{2}\Pi$ are all the allowed 
ones of $J_{A}$$-$$J_{5\Sigma}$=0,$\pm$1.
According to our (limited) checks, this initial-final rotational selectivity is insignificant to the results.


The theoretical results for the chirp dependence of $I_{5p}$  are presented in Fig.~\ref{fig:exp_theo_results} (black lines). 
As seen, they reproduce the experimental ones to a very high degree, both qualitatively and quantitatively.
Qualitatively, they exhibit all 
the experimental features of enhancement by negative chirps  
as well as the experimental trend of the general chirp dependence.
The former includes the larger-than-1 $\mathsf{EF_{N/P}}(|k|)$ 
for all the $|k|$ values, 
with the higher $\mathsf{EF_{N/P}}$ values obtained beyond small $|k|$ values with a weak chirp dependence,
and the larger-than-1 $\mathsf{EF}(k)$ for small negative chirps $k$.
Quantitatively, the theory-experiment fit of the $\mathsf{EF}$ results is excellent  
over the chirp range of $-$0.02 to 0.002 ps$^{2}$/rad and above 0.02~ps$^{2}$/rad,
including the corresponding maximal $\mathsf{EF}$ value of about 1.10, 
while at the other chirp values there is a small deviation of theory from experiment. 
This small deviation leads to calculated $\mathsf{EF_{N/P}}$ values that are somewhat  
higher than the experimental ones.
We associate this theoretical deviation in the $\mathsf{EF}$ values    
with the spin-orbit coupling not included in our KAr model. 


Following the theory-experiment agreement, we further 
analyze the system to identify the chirp control mechanism.
To this end, we compare
the spectral locations of the two excitation Franck-Condon windows (FCWs),
where the pulse spectrum is resonant with the corresponding potential difference. 
As seen in Fig.~\ref{fig:pot_exc_scheme}, for the X$^{2}\Sigma^{+}$$-$A$^{2}\Pi$ FCW it is the full pulse spectrum, 
while for the A$^{2}\Pi$$-$$(5)^{2}\Sigma^{+}$ FCW it is only the low-frequency edge of the spectrum.
Hence, most of the $(5)^{2}\Sigma^{+}$'s population is excited via intermediates 
for which the initial-to-intermediate transition frequency ($\omega_{ni}$) is higher than the intermediate-to-final     
one ($\omega_{fn}$), i.e., $\omega_{ni}$$>$$\omega_{fn}$.
These dominant intermediate eigenstates are energetically connected  
by the pulse spectrum to both the initial and final eigenstates and 
have significant VTDMs with both. 
As previously shown 
for a (bound) three-level (initial-intermediate-final) system~\cite{ChatelPRA03-ResMed1plus1-chirp,ChatelPRAA04-ResMed1plus1-chirp}, 
a weak-field resonance-mediated two-photon excitation via such an intermediate (for which $\omega_{ni}$$>$$\omega_{fn}$)
is enhanced by negative chirps.
From a time-domain viewpoint, 
the negatively-chirped pulses photo-induce an optimal excitation sequence: 
the temporally-decreasing instantaneous pulse frequency  
comes into resonance first with $\omega_{ni}$ and only then with $\omega_{fn}$. 
From a frequency-domain viewpoint,  
the negative chirps 
enhance the degree of constructive interferences, versus destructive ones, 
among the two-photon transition pathways.
Therefore, we identify the present control mechanism to be the combination of  
(i) Franck-Condon filtering~\cite{RybakPRL11,AmaranJCP13} 
comprised of two steps, which purifies 
the vast thermal ensemble of excitation channels and selects 
a dominant sub-ensemble of channels that exhibit a uniform susceptibility to chirp control,  
and 
(ii) chirp-dependent resonance-mediated coherent excitation of each of the selected channels. 
The overall enhancement by negative chirps is enabled here, 
even though the number of selected channels is still very large,  
since they are all individually characterized by transition frequencies for which negative chirps are optimal.

The crucial role of the two-step Franck-Condon filtering (TS-FCF) 
is also illustrated  
by artificially setting in the calculations 
all the system's VTDMs to the same value. 
This setting eliminates the TS-FCF, 
such that the  
excitation (under the electronic and rotational selection rules) is affected only by the state energies and the pulse spectrum.
As seen in Fig.~\ref{fig:exp_theo_results} (gray lines), without the TS-FCF 
the enhancement by negative chirps is almost completely gone,
with all the $\mathsf{EF_{N/P}}$ values being about~1 and the maximal signal obtained for the TL pulse.


\begin{SCfigure*}
\includegraphics[width=0.69\textwidth]{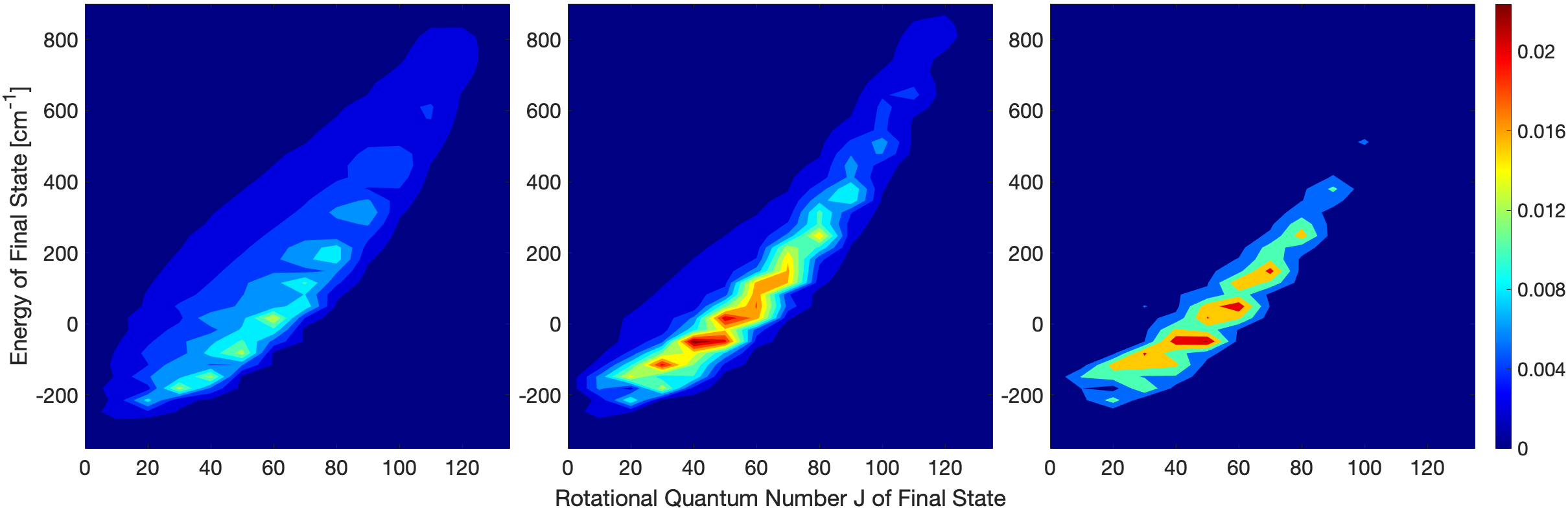}
\caption{Theoretical results for coherent control over the final population distribution.
The relative population of the final eigenstates in $(5)^{2}\Sigma^{+}$, 
as a function of their rotational quantum number J and energy,
is shown for the TL pulse (left panel) and linearly-chirped pulses of chirps 
 $-$0.01 (middle panel) and $-$0.02~rad/ps$^{2}$ (right panel).
Each distribution is normalized to a population sum of 1.}
\label{fig:theo_final_pop_results}
\end{SCfigure*}


Last, we theoretically analyze the final population distribution and its chirp control.
The distribution of the relative population of the different final eigenstates in $(5)^{2}\Sigma^{+}$ 
is shown in Fig.~\ref{fig:theo_final_pop_results}
following, for example, excitations by the TL pulse and linearly-chirped pulses of chirps $-$0.01 and $-$0.02~rad/ps$^{2}$.  
The population is presented as a function of the eigenstate's rotational quantum number J$_{5\Sigma}$  and 
energy $\mathcal{E}_{5\Sigma}$ (relative to the $(5)^{2}\Sigma^{+}$ asymptote). 
Each distribution is normalized to a population sum~of~1.
In terms of the general characteristics of the distribution, the following is observed for all the pulses.
The excitation involves high rotations of up to J$_{5\Sigma}$$\sim$130, 
with the maximal population being excited at J$_{5\Sigma}$$\sim$40$-$70. 
The main distribution feature is that 
the excited population shows a clear correlation between J$_{5\Sigma}$ and $\mathcal{E}_{5\Sigma}$:
there is only a selected 
band of $\mathcal{E}_{5\Sigma}$ that gets populated for a given J$_{5\Sigma}$, 
and these energies increase as J$_{5\Sigma}$ increases.
This correlation originates from the first step of the TS-FCF,
where, for an initial rotation J$_{X}$, there is only a selected band of collision energy $\mathcal{E}_{X}$
that allows the collision pair to properly reach the X$-$A FCW at the 
short internuclear distances.
Only at these collision energies, the scattering state's wavefunction has large enough non-oscillatory amplitude 
within this FCW. 
Such 
rotational-translational correlation existing for the initial eigenstates of the dominant excitation channels 
directly leads to a similar correlation for their final eigenstates.
This is because,  
from a given initial energy $\mathcal{E}_{X}$, there is only a limited range of final energies $\mathcal{E}_{5\Sigma}$ 
that are effectively accessible by the weak-field two-photon excitation with a given pulse spectrum, 
and the location of this final range increases as $\mathcal{E}_{X}$ increases.
In terms of the coherent control,  
there is a clear and prominent chirp dependence of the final population distribution 
seen for negative chirps:
as the negative chirp magnitude increases from zero chirp, 
the distribution gets narrower in terms of both J$_{5\Sigma}$ and $\mathcal{E}_{5\Sigma}$, 
with an increased correlation between J$_{5\Sigma}$ and $\mathcal{E}_{5\Sigma}$.
For all the present positive chirps (not shown here), 
the distribution shape is almost chirp independent and stays very close to the one excited by the~TL~pulse.


In summary, we experimentally and theoretically demonstrate for the first time weak-field coherent control of ultrafast molecule making,
achieving it even under thermally hot conditions typical of chemical reactions.
It is demonstrated with hot pairs of colliding K and Ar atoms using linearly-chirped femtosecond pulses.
Our {\it ab initio} modeling has identified the control mechanism to combines 
Franck-Condon filtering and 
phase-dependent 
shaped excitation of the selected excitation channels.
The former purifies the huge 
thermal ensemble of excitation channels and selects 
a sub-ensemble of channels  
that is 
susceptible for coherent control, 
while the latter realizes their control  
by manipulating intra-channel 
interferences. 
The demonstrated control of the KAr system serves as a basic model for 
controlling triatomic systems,
with an atom and a diatomic molecule as the reactants, 
that are non-reactive in their ground electronic state and all their excited electronic states are weakly-bound or unbound. 
Our experiments are also the first to achieve weak-field femtosecond coherent control of any initially-unbound~system.
With the weak-field feasibility, 
the toolbox for coherent control of ultrafast molecule making 
can now handle many new excitation scenarios that  
their control is superior or only possible with weak fields.
This opens~up
new feasible routes for the coherent control of binary chemical photoreactions.

Financial support from the Deutsche Forschungsgemeinschaft (DFG—German Research Foundation) 
under the DFG Priority Program 1840, 
‘Quantum Dynamics in Tailored Intense Fields',  
is gratefully acknowledged.

\bibliography{KAr}

\end{document}


\title{Supplemental Material: Weak-Field Coherent Control of Ultrafast Molecule Making}


\author{Moran Geva}
\affiliation{The Shirlee Jacobs Femtosecond Laser Research Laboratory, Schulich Faculty of Chemistry, Technion-Israel Institute of Technology, Haifa 32000, Israel}

\author{Yonathan Langbeheim}
\affiliation{The Shirlee Jacobs Femtosecond Laser Research Laboratory, Schulich Faculty of Chemistry, Technion-Israel Institute of Technology, Haifa 32000, Israel}

\author{Arie Landau}
\affiliation{Institute of Advanced Studies in Theoretical Chemistry and Schulich Faculty of Chemistry, Technion-Israel Institute of Technology, Haifa 32000, Israel}

\author{Zohar Amitay}
\affiliation{The Shirlee Jacobs Femtosecond Laser Research Laboratory, Schulich Faculty of Chemistry, Technion-Israel Institute of Technology, Haifa 32000, Israel}

 

\maketitle 

\section{\hspace{1.75cm} Frequency-Domain Perturbative Description \newline of the Present Weak-Field Ultrafast Molecule Making}

The final 
excitation amplitude $A_{f}^{(i \rightarrow f)}$ 
of a given (bound or unbound) final 
eigenstate $\left| f \right>$ in $(5)^{2}\Sigma^{+}$ 
that is photo-induced by the weak  
resonance-mediated two-photon femtosecond excitation 
from a given initial 
(unbound) 
eigenstate $\left| i \right>$ in X$^{2}\Sigma^{+}$ 
via a manifold of $N$ (bound and unbound) 
intermediate eigenstates $\left| n \right>$ ($n=1..N$) 
in A$^{2}\Pi$ 
is obtained using second-order  
time-dependent perturbation theory. 
It is an extension of the three-level case 
with a single intermediate~\cite{SilberbergPRL01-ResMed1plus1,ChatelPRA03-ResMed1plus1-chirp,SilberbergPRL04-ResMed1plus1-PolShaping,ChatelPRAA04-ResMed1plus1-chirp}
to the present case with multiple intermediates.
With a linearly-polarized shaped femtosecond pulse, it is given in the frequency domain by 
\begin{equation}
A_{f}^{(i \rightarrow f)}  =  \SumInt_{n=1}^{N} A_{f}^{(i \rightarrow n \rightarrow f)}  =  
- \frac{1}{i \hbar^2} 
\SumInt_{n=1}^{N} \Bigg\{ \Bigg.  \mu_{fn} \mu_{ni} 
\bigg[ \bigg.  i  \pi E(\omega_{ni}) E(\omega_{fi}-\omega_{ni})   
+ \wp\int_{-\infty}^{\infty} \frac{1}{\omega_{ni}-\omega} E(\omega) E(\omega_{fi}-\omega) d\omega \bigg] \Bigg\} \; ,
\label{eq_A}
\end{equation}
where 
$\SumInt_{n}$ stands for the summation over the intermediate discrete bound eigenstates and 
proper integral over the intermediate continuum unbound eigenstates, 
and $\wp$ is the Cauchy principal value. 
The $\omega_{kl} = (\mathcal{E}_{k}-\mathcal{E}_{l})/\hbar$ is the resonance frequency between a pair of eigenstates,
with $\mathcal{E}_{k}$ and $\mathcal{E}_{l}$ being their energies, 
and $\mu_{kl}$ is the component of the transition dipole moment between them along the 
linear-polarization axis of the pulse.
The energies of the initial state $\left| i \right>$, the final state $\left| f \right>$ and a given intermediate state $\left| n \right>$ 
are $\mathcal{E}_{i}$, $\mathcal{E}_{f}$ and $\mathcal{E}_{n}$, respectively,
with $\omega_{fi} = \omega_{fn} + \omega_{ni}$.
The $E(\omega)=\left|E(\omega)\right| \exp(-i\Phi(\omega))$ is the broad spectral field of the pulse, 
with $\left|E(\omega)\right|$  and $\Phi(\omega)$ being the spectral amplitude and phase, respectively. 
As seen, the amplitude $A_{f}^{(i \rightarrow f)}$ results from a coherent summation over all the 
different amplitudes $A_{f}^{(i \rightarrow n \rightarrow f)}$, each originating  
from the excitation via a different single intermediate eigenstate $\left| n \right>$.
In turn, the $A_{f}^{(i \rightarrow n \rightarrow f)}$ corresponding to a given $\left| n \right>$ 
results from a coherent summation over all the 
different amplitudes that are contributed by all the different initial-to-final resonance-mediated two-photon excitation pathways via $\left| n \right>$.
Its first term includes the on-resonant pathway (with regard to $\left| n \right>$), while its second term incorporates all the near-resonant pathways.
The corresponding excitation probability $P_{f}^{(i \rightarrow f)}$ of $\left| f \right>$ 
is given by 
\begin{equation}
P_{f}^{(i \rightarrow f)} =  \left| A_{f}^{(i \rightarrow f)} \right|^{2} \; .
\label{eq_P}
\end{equation}


Resulting from all the separate excitation channels that simultanously occur, 
the total final population 
$\mathsf{P}_{f}$ 
of $\left| f \right>$ 
is the outcome of an incoherent integral  
over contributions $P_{f}^{(i \rightarrow f)}$ from all the different initial unbound scattering eigenstates $\left| i \right>$ in X$^{2}\Sigma^{+}$ 
weighted according to their initial thermal population.
Ignoring the spin, an unbound eigenstate in X$^{2}\Sigma^{+}$,
$\left| i \right> \equiv \left| {\rm X}^{2}\Sigma^{+}\!\!, J_{\rm X}, M_{\rm X}, \Omega_{\rm X}\!=\!0, \mathcal{E}_{\rm X} \right> $, 
is characterized by its rotational quantum numbers 
$\left\{ J_{\rm X}, M_{\rm X}, \Omega_{\rm X}\!=\!0 \right\}$ 
and by its continuum energy  (relative the X-state asymptote)  $\mathcal{E}_{\rm X}$.
The former  
correspond, respectively, to the 
total orbital angular momentum of the molecule,
its projection on the pulse's linear-polarization axis and its zero projection on the internuclear axis. 
Hence, $\mathsf{P}_{f}$ is 
given by
\begin{equation}
\mathsf{P}_{f} = 
\frac{1}{\mathcal{Z}} 
\sum_{J_{X},M_{X}}  
\int_{0}^{\infty} d\mathcal{E}_{\rm X} \, \rho(\mathcal{E}_{\rm X} ; J_{X}) \, \exp(- \frac{\mathcal{E}_{\rm X}}{k_{_B}T}) \, 
P_{f}^{\left( \left\{ {\rm X}^{2}\Sigma^{+}\!\!, J_{\rm X}, M_{\rm X}, \Omega_{\rm X}=0, \mathcal{E}_{\rm X} \right\} \rightarrow  f \right)} \; ,
\label{eq3}
\end{equation}
where ${\mathcal{Z}}$ is the partition function, 
$\rho(\mathcal{E}_{\rm X} ; J_{X})$ is the density of unbound (continuum) states of X$^{2}\Sigma^{+}$ at energy $\mathcal{E}_{\rm X}$ for a given $J_{X}$, 
and $T$ is the temperature of the initial thermally-hot ensemble of colliding K and Ar atoms. 

The components of the transition dipole moments along the linear-polarization axis of the pulse (Z direction),
which are incorporated in Eq.~(\ref{eq_A}), are given as follows.
Similar to 
an eigenstate in X$^{2}\Sigma^{+}$, 
an eigenstate in A$^{2}\Pi$ is defined here as 
$\left| n \right> \equiv \left| {\rm A}^{2}\Pi, J_{\rm A}, M_{\rm A}, \Omega_{\rm A}\!=\!1, \left\{ \nu_{\rm A} {\rm \; or \;} \mathcal{E}_{\rm A} \right\} \right>$, 
where 
bound and unbound eigenstates are characterized by 
their vibrational quantum number $\nu_{\rm A}$  
and continuum energy (relative to the A-state asymptote)~$\mathcal{E}_{\rm A}$, respectively.
So, the initial-intermediate transition-dipole-moment Z-component $\mu_{ni} $ is given by 
\begin{eqnarray}
\mu_{ni} & = & \mu_{ni}^{(VTDM)} \cdot \mu_{ni}^{(ROT, \, \Sigma\rightarrow\Pi, \,\Delta{M}=0, \,\Delta{J}=0,\pm1)} \; , \nonumber \\ 
\nonumber \\ \nonumber
\mu_{ni}^{(VTDM)} & = & 
\left< {\rm A}^{2}\Pi, J_{\rm A}, \Omega_{\rm A}\!=\!1, \left\{ \nu_{\rm A} {\rm \; or \;} \mathcal{E}_{\rm A} \right\} 
\left| \mu_{{\rm A}^{2}\Pi,{\rm X}^{2}\Sigma^{+}}^{(ETDM)}(R) \right|  
{\rm X}^{2}\Sigma^{+}\!, J_{\rm X}, \Omega_{\rm X}\!=\!0, \mathcal{E}_{\rm X}  \right>  \; , \nonumber \\
\nonumber \\ \nonumber
\mu_{ni}^{(ROT, \, \Sigma\rightarrow\Pi, \,\Delta{M}=0, \,\Delta{J}=0,\pm1)} & = & 
\nonumber
(-1)^{(M_{X}-1)} \cdot \sqrt{(2J_{X}+1)(2J_{A}+1)} \cdot
\begin{pmatrix}
 J_{X}   &  1  &  J_{A}  \\
-M_{X}  &  0  &  M_{A}
\end{pmatrix}
\begin{pmatrix}
    J_{X}                  &   1       &  J_{A}  \\
           0                 &  -1       &  1
\end{pmatrix}
\; , \nonumber
\end{eqnarray}
where 
$\mu_{ni}^{(VTDM)}$ is the corresponding vibrationally-averaged transition dipole moment (VTDM)
and 
$\mu_{ni}^{(ROT, \, \Sigma-\Pi, \,\Delta{M}=0, \,\Delta{J}=0,\pm1)}$ is the corresponding rotational transition dipole moment component
describing a one-photon $\Sigma$-to-$\Pi$ molecular transition 
with angular-momentum selection rules of $\Delta{M}$=0 and $\Delta{J}$=0,$\pm$1 
(expressed using 3-j symbols)~\cite{ZareBook}.
The $\mu_{{\rm A}^{2}\Pi,{\rm X}^{2}\Sigma^{+}}^{(ETDM)}(R)$ is the electronic transition dipole moment (ETDM) between the X$^{2}\Sigma^{+}$ and A$^{2}\Pi$ states. 
Likewise, an eigenstate in $(5)^{2}\Sigma^{+}$ is given here by 
$\left| f \right> \equiv \left| (5)^{2}\Sigma^{+}\!,\!J_{5\Sigma}, M_{5\Sigma}, \Omega_{5\Sigma}\!=\!0, \left\{ \nu_{5\Sigma} {\rm \; or \;} \mathcal{E}_{5\Sigma} \right\} \right>$
and the intermediate-final transition-dipole-moment Z-component $\mu_{fn} $ is given by 
\begin{eqnarray}
\mu_{fn} & = & \mu_{fn}^{(VTDM)} \cdot \mu_{fn}^{(ROT, \, \Pi\rightarrow\Sigma, \,\Delta{M}=0, \,\Delta{J}=0,\pm1)} \; , \nonumber \\ 
\nonumber \\ \nonumber
\mu_{fn}^{(VTDM)} & = & \left< (5)^{2}\Sigma^{+}\!, J_{5\Sigma}, \Omega_{5\Sigma}\!=\!0,\left\{ \nu_{5\Sigma} {\rm \; or \;} \mathcal{E}_{5\Sigma} \right\}
\left| \mu_{(5)^{2}\Sigma^{+}\!,{\rm A}^{2}\Pi}^{(ETDM)}(R) 
\right| {\rm A}^{2}\Pi, J_{\rm A}, \Omega_{\rm A}\!=\!1, \left\{ \nu_{\rm A} {\rm \; or \;} \mathcal{E}_{\rm A} \right\} \right> \; \nonumber \\
\nonumber \\ \nonumber
\mu_{fn}^{(ROT, \, \Pi\rightarrow\Sigma, \,\Delta{M}=0, \,\Delta{J}=0,\pm1)} & = & 
(-1)^{M_{A}}  \cdot \sqrt{(2J_{X}+1)(2J_{A}+1)} \cdot
\begin{pmatrix}
 J_{A}   &  1  &  J_{5\Sigma}  \\
-M_{A}  &  0  &  M_{5\Sigma}
\end{pmatrix}
\begin{pmatrix}
    J_{A}                   &   1        &  J_{5\Sigma}  \\
         -1                   &   1        &  0
\end{pmatrix}
\; . \nonumber
\end{eqnarray}

\bibliography{KAr}